\documentclass[twocolumn,twocolappendix]{aastex702}

\usepackage{graphicx}
\usepackage{subcaption}
\usepackage{txfonts}
\usepackage{longtable}
\usepackage{subcaption}
\usepackage{float} 
\usepackage{orcidlink}
\usepackage{threeparttable}
\usepackage{array}
\usepackage{caption}
\usepackage{makecell}

\begin{document}

\title{
Searching for embedded protoplanets with the Keck/NIRC2 Vortex Coronagraph:\\
Confirmation of a core-accretion planet in the narrow gap of the Elias 2-24 disk
}

\author[orcid=0009-0001-3428-1029,sname='Bernardi']{Andrea Bernardi}
\affiliation{Instituto de Estudios Astrof\'isicos, Facultad de Ingenier\'ia y Ciencias, Universidad Diego Portales, Av. Ej\'ercito 441, Santiago, Chile}
\affiliation{Millennium Nucleus on Young Exoplanets and their Moons (YEMS), Santiago, Chile}
\email[show]{andrea.bernardi@mail.udp.cl}  

\author[orcid=0000-0002-5903-8316]{Alice Zurlo}
\affiliation{Instituto de Estudios Astrof\'isicos, Facultad de Ingenier\'ia y Ciencias, Universidad Diego Portales, Av. Ej\'ercito 441, Santiago, Chile}
\affiliation{Millennium Nucleus on Young Exoplanets and their Moons (YEMS), Santiago, Chile}
\email{alice.zurlo@mail.udp.cl}  

\author[orcid=0000-0002-2828-1153]{Lucas A. Cieza}
\affiliation{Instituto de Estudios Astrof\'isicos, Facultad de Ingenier\'ia y Ciencias, Universidad Diego Portales, Av. Ej\'ercito 441, Santiago, Chile}
\affiliation{Millennium Nucleus on Young Exoplanets and their Moons (YEMS), Santiago, Chile}
\email{lucas.cieza@mail.udp.cl}  

\author[orcid=0000-0003-4769-1665]{Garreth Ruane}
\affiliation{Jet Propulsion Laboratory, California Institute of Technology, 4800 Oak Grove Dr, Pasadena, CA, 91109, USA}
\email{garreth.ruane@jpl.nasa.gov}  

\author[orcid=0000-0002-0101-8814]{Valentin Christiaens}
\affiliation{Institute of Astronomy, KU Leuven, Celestijnenlaan 200D, Leuven, Belgium}
\affiliation{STAR Institute, Université de Liège, Allée du Six Août 19c, 4000 Liège, Belgium}
\email{valentin.christiaens@uliege.be} 

\author[orcid=0009-0009-8115-8910]{Anuroop Dasgupta}
\affiliation{Instituto de Estudios Astrof\'isicos, Facultad de Ingenier\'ia y Ciencias, Universidad Diego Portales, Av. Ej\'ercito 441, Santiago, Chile}
\affiliation{Millennium Nucleus on Young Exoplanets and their Moons (YEMS), Santiago, Chile}
\affiliation{European Southern Observatory, Alonso de Córdova 3107, Vitacura, Santiago, Chile}
\email{anuroopdasgupta1999@gmail.com} 

\author[orcid=0000-0002-7002-8928]{Greta Guidi}
\affiliation{University Grenoble Alpes, CNRS, IPAG, Grenoble, France}
\email{greta.guidi@univ-grenoble-alpes.fr} 

\author[orcid=0000-0002-8895-4735]{Dimitri Mawet}
\affiliation{Department of Astronomy, California Institute of Technology, Pasadena, CA, USA}
\affiliation{Jet Propulsion Laboratory, California Institute of Technology, 4800 Oak Grove Dr, Pasadena, CA, 91109, USA}
\email{dmawet@astro.caltech.edu} 

\author[orcid=0000-0001-8467-1933]{Dino Mesa}
\affiliation{INAF-Osservatorio Astronomico di Padova, Vicolo dell'Osservatorio 5, Padova, Italy, 35122-I}
\email{dino.mesa@inaf.it} 

\author[orcid=0000-0003-2953-755X]{Sebastián Pérez}
\affiliation{Millennium Nucleus on Young Exoplanets and their Moons (YEMS), Santiago, Chile}
\affiliation{Departamento de Física, Universidad de Santiago de Chile, Av. Victor Jara 3659, Santiago, Chile}
\affiliation{Center for Interdisciplinary Research in Astrophysics and Space Science (CIRAS), Universidad de Santiago de Chile, Chile}
\email{sebastian.astrophysics@gmail.com} 

\author[orcid=0000-0001-5058-695X]{Jonathan P. Williams}
\affiliation{Institute for Astronomy, 2680 Woodlawn Dr, Honolulu, HI, USA}
\email{jw@hawaii.edu} 

\begin{abstract}

We present high-contrast imaging observations of seven stars obtained with the Keck/NIRC2 vortex coronagraph between May and June 2018 in the $L'$ and $M_s$ bands. All targets host protoplanetary disks with substructures such as gaps and cavities, potentially tracing ongoing planet formation. Our aim is to assess whether these embedded planets can be directly detected. We reduced the observations with a custom pipeline based on the \texttt{VIP} Python package. Post-processed images were obtained using angular differential imaging and reference differential imaging combined with principal component analysis. Finally, $5\sigma$ contrast curves and detection limits were derived for all targets. We confirm the planetary candidate around Elias 2-24, previously proposed in the literature, at a separation of $394\pm31$ mas ($54.9\pm4.3$ au) and a position angle of $298.8\pm3.2^\circ$, within the narrow gap of the protoplanetary disk. Comparing the derived photometry with 1 Myr isochrones, we infer a mass between 1.9 and 4.0 M\textsubscript{Jup} for Elias 2-24 b, without considering accretion effects, in agreement with estimates considering the properties of the gap and planet-disk interactions. For all the systems in the survey, the $5\sigma$ detection limits reach sensitivities down to 1-11 M\textsubscript{Jup} at separations $\gtrsim$ 0.25$^{\prime\prime}$. Since Elias 2-24 b (age $\lesssim$ 1 Myr) is the youngest exoplanet discovered to date, its confirmation has important implications for planet formation timescales and the origin of the gaps observed in very young disks.

\end{abstract}



\section{Introduction} 
\label{sec:intro}
The discovery and characterization of young giant planets embedded in protoplanetary disks is fundamental for understanding planet formation mechanisms. While core accretion \citep{Pollack1996} can proceed in a wide range of protoplanetary disks, gravitational instability \citep{Boss1997} is thought to operate only in the most massive and unstable systems, where self-gravity drives spiral arms that may subsequently fragment into bound companions \citep{Zhu2012,Weber2025,Dasgupta2025}. 

Thanks to the advent of new-generation 8–10 m telescopes and advanced adaptive optics systems, these companions can now be directly detected through high-contrast imaging at separations approximately larger than 10 au. Examples include the accreting protoplanets around PDS 70 \citep{Keppler2018, Mesa2019, Haffert2019}, the two planets around WISPIT 2 \citep{vanCapelleveen2025, Close2025, Lawlor2026}, and HD 169142 b \citep{Hammond2023}.

A different method for detecting such companions has become possible thanks to the capabilities of the Atacama Large Millimetre/submillimetre Array (ALMA). Its sensitivity enables high-resolution mapping of disk kinematic structures, potentially leading to the discovery of otherwise unseen companions (e.g., \citealt{Perez2018, Garg2022, Pinte2023a} for a review). In this context, large ALMA surveys aimed at studying protoplanetary disks (e.g., \citealt{Teague2025}) are crucial for investigating disk evolution, the possible presence of companions, and the effects of these companions on disk properties.

\begin{deluxetable*}{lccccccccc}
\tablecaption{Keck/NIRC2 Vortex observations. $L'$ magnitudes are estimated via logarithmic interpolation between WISE W1 and W2 bands. The $M_s$ magnitude of PDS 70 is approximated using the WISE W2 band.\label{obs}}
\label{obs}
\tablewidth{\textwidth}
\tabletypesize{\footnotesize}
\tablehead{
\colhead{Name} & \colhead{Distance} & \colhead{Age} & \colhead{Spectral type} & \colhead{Obs date} & \colhead{$\Delta$ PA} & \colhead{Seeing} & \colhead{t\textsubscript{exp}} & \colhead{\shortstack{Apparent\\magnitude}} & \colhead{References} \\
\colhead{} & \colhead{(pc)} & \colhead{(Myr)} & \colhead{} & \colhead{(UT)} & \colhead{($^{\circ}$)} & \colhead{('')} & \colhead{(s)} & \colhead{(mag)} & \colhead{}
}
\startdata
\multicolumn{10}{c}{$L'$ observations} \\
\hline
V1094 Sco & $154.76\pm0.76$ & $2.5^{+2.0}_{-1.0}$ & K6 & 2018-05-25 & $30.21$ & $0.45$ & $3510$ & $7.802\pm0.023$ & 1, 2, 3, 4 \\[0.75ex]
AS 209 & $121.25\pm0.43$ & 1-2 & K4V & 2018-05-26 & $26.97$ & $0.59$ & $2700$ & $6.487\pm0.036$ & 1, 5, 6, 4 \\[0.75ex]
HD 98800 B & $44.9\pm4.7$ & $\sim10$ & K5 & 2018-05-27 & $28.36$ & $0.47$ & $2880$ & $5.478\pm0.060$ & 7, 8, 9, 4 \\[0.75ex]
WSB 82 & $145.77\pm0.86$ & 2 & mid K & 2018-06-05 & $38.39$ & $1.02$ & $4185$ & $6.363\pm0.040$ & 1, 10, 11, 4 \\[0.75ex]
Elias 2-24 & $139.3\pm1.2$ & $\lesssim1$ & K6 & 2018-06-21 & $44.84$ & $0.50$ & $3330$ & $6.439\pm0.041$ & 1, 12, 13, 4 \\[0.75ex]
RX J1633.9-2442 & $143.84\pm0.71$ & 2 & K7 & 2018-06-21 & $6.13$ & $0.50$ & $810$ & $8.720\pm0.025$ & 1, 10, 14, 4 \\[0.75ex]
\hline
\multicolumn{10}{c}{$M_s$ observations} \\
\hline
PDS 70 & $112.39\pm0.24$ & $5.4\pm1.0$ & K7IV & 2018-06-05 & $22.86$ & $1.02$ & $2700$ & $7.714\pm0.020$ & 1, 15, 16, 4 \\[0.75ex]
\enddata

\tablecomments{(1) \citet{GaiaDR3}; (2) \citet{Garufi2020}; (3) \citet{Alcala2017}; (4) \citet{Cutri2012}; (5) \citet{Oberg2021}; (6) \citet{Torres2006};  (7) \citet{vanLeeuwen2007}; (8) \citet{Torres2008}; (9) \citet{Zurlo2020}; (10) \citet{Cieza2019}; (11) \citet{Esplin2018};  (12) \citet{Andrews2018}; (13) \citet{Ricci2010}; (14) \citet{Orellana2012};    (15) \citet{Muller2018}; (16) \citet{Pecaut2016}.}
\end{deluxetable*}

In this work, we present a Keck/NIRC2 vortex coronagraph \citep{Serabyn2017} survey of stars hosting protoplanetary disks previously observed at other wavelengths, showing substructures such as gaps and/or cavities potentially indicative of embedded protoplanets. With our observations, we confirm the presence of a Jovian planet in the disk of Elias 2-24, consistent with the detections reported by \citet{Jorquera2021} and \citet{Pinte2023b}. 

The Letter is organized as follows. Section~\ref{sec:obs_red} describes the sample selection as well as the observing and data reduction procedures. In Section~\ref{sec:results} we report our results. Section \ref{sec:characterization} describes the characterization of Elias 2-24 b, which is further discussed in Section~\ref{sec:discussion}. The conclusions are presented in Section~\ref{sec:conclusion}.

\section{Observations and data reduction}
\label{sec:obs_red}

\subsection{Sample presentation}
Our sample was initially drawn from the targets included in the ODISEA survey \citep{Cieza2019}. More specifically, we focused on young ($\lesssim$10 Myr) systems hosting disks with clear substructures in millimeter/submillimeter, such as gaps, rings, and cavities, potentially carved by embedded protoplanets. This yielded eight potential targets, including Elias 2-24, WSB 82, and RX J1633.9-2442. We also selected several targets not included in the ODISEA survey but meeting the same selection criteria, which served as backup targets in case any of the primary ODISEA targets could not be observed. In particular, AS 209 was part of the DSHARP survey \citep{Andrews2018, Guzman2018}, showing multiple rings and bright emission features. ALMA observations of V1094 Sco \citep{vanTerwisga2018} revealed a very extended disk with two gap-ring pairs and continuum emission out to 300 au. HD 98800 is a quadruple system of two spectroscopic binaries \citep{Torres1995,Zurlo2020}, its BaBb component hosts a circumbinary disk \citep{Skinner1992}, perpendicular to the binary orbital plane \citep{Kennedy2019}. Finally, PDS 70 hosts a disk with a large gap detected first in the near infrared \citep{Hashimoto2012}, and then confirmed by ALMA \citep{Long2018}, within which at least two accreting protoplanets have been discovered \citep{Keppler2018, Mesa2019, Haffert2019,Hammond2025}, as well as a dusty circumplanetary disk around one of the companions \citep{Isella2019}.

\subsection{Observing strategy}

The sample of seven targets was observed between 25 May and 21 June 2018 (program H284, PI J.P. Williams). Further details on the sample properties and observations are presented in Table \ref{obs}. All observations were obtained with the vector vortex coronagraph installed in Keck/NIRC2, using the \texttt{QACITS} real-time coronagraphic PSF-centering algorithm \citep{Huby2015, Huby2017}. All epochs are presented here for the first time, except for the 26 May 2018 observation of AS 209, previously presented in \citet{Wallack2024}. The typical centering accuracy of \texttt{QACITS} is 2.4 mas rms \citep{Huby2017}, corresponding to $\approx0.025\,\lambda/D$ rms in the $L'$ band. For comparison, the pixel scale of the NIRC2 vortex coronagraph is 9.971 mas pixel$^{-1}$ \citep{Service2016}. All targets, except for PDS 70, were observed in the $L'$ band (central wavelength $3.776\,\mu$m). PDS 70 was observed in the $M_s$ band (central wavelength $4.670\,\mu$m) to estimate the $L$--$M$ color of its protoplanets in the event of a detection (more details in Appendix~\ref{app:pds70}).

The median integration time is 48.0 minutes, and the median seeing, as measured by the Maunakea Weather Center DIMM Seeing Monitor\footnote{\url{http://mkwc.ifa.hawaii.edu/current/seeing/}}, was $0.5^{\prime\prime}$. Each observing sequence consisted of an unocculted image of the star to characterize the PSF, a sky frame obtained on a blank field $10^{\prime\prime}$ from the target, and 10–30 science frames with the star centered on the vortex, each with an integration time of 45 s, obtained from the co-addition of 45 frames with an integration time of one second (except for PDS 70, for which 90 frames, each with an integration time of 0.5 s, were co-added). For longer observations or rapidly varying conditions, the sequence was repeated every 10–30 minutes to monitor variations in the PSF and sky background. All observations were obtained in pupil-tracking mode close to the meridian passage, enabling angular differential imaging (ADI, \citealt{Marois2006}) processing. The median parallactic angle rotation of the sample is $28.36^{\circ}$.

\subsection{Data reduction}
The raw data were processed with a custom pipeline based on the Vortex Image Processing package (\texttt{VIP}; \citealt{GomezGonzalez2017, Christiaens2023}). We first applied dark-field and flat-field corrections to the science and sky frames. Using \texttt{VIP}, bad, hot, and dead pixels were replaced with the median value within a $5\times5$ neighboring box. We then corrected for geometrical distortion using the solution from \citet{Service2016} and removed the sky background with a principal component analysis (PCA)-based algorithm. Finally, the sky-subtracted images were registered to the stellar position by aligning the speckle pattern with the median frame through cross-correlation. We then removed the stellar point spread function (PSF) using a principal PCA-ADI approach implemented in \texttt{VIP}, exploring different numbers of principal components (PCs) together with both annular and full-frame PCA methods. For the annular PCA, we adopted a 0.2-Full Width at Half Maximum (FWHM) parallactic angle threshold (to partially mitigate the effects of self-subtraction), except for RX J1633.9-2442, where the limited field rotation prevented its use, together with a 1-FWHM inner mask and 15-pixel (0.15$^{\prime\prime}$) wide annuli. As PSF models, we used the unocculted stellar PSF images of each target after background subtraction and rejection of low-quality frames. Astrometry and photometry for all detected point sources were derived using the NEGFC technique \citep{Marois2010, Lagrange2010} implemented in \texttt{VIP}, yielding the companion separation, position angle, and flux ratio relative to the host star. Initial estimates of the position and flux were obtained with the \texttt{firstguess()} function and refined using \texttt{mcmc\_negfc\_sampling()}. The astrometric uncertainties include contributions from the plate scale error, residual distortion, True North alignment, and the Monte Carlo Markov Chain (MCMC) fitting procedure. The final uncertainties were estimated following the method described in \citet{Franson2022}.

\subsubsection{Disks retrieval}
\label{subsec:tentativedisks}
To investigate the detectability of circumstellar disks in our observations, we applied a reference differential imaging (RDI) approach. Since constructing a same-night reference library was not possible, we used all the $L'$-band observations in the survey as references. To optimize the frame selection and improve small-separation contrasts, we ranked the frames using the mean-squared error (MSE) and Pearson correlation coefficient (PCC). The optimal number of reference frames was determined by injecting a fake planet into the science cubes and measuring its signal-to-noise ratio after PCA-RDI while iterating over the library size. Further details about this technique are provided by \citet{Ruane2019}.

We also applied an alternative technique, better suited to cases where standard RDI is not feasible or provides poor-quality results. The method consists of de-rotating the science cube, computing its median image, and applying a high-pass filter by subtracting a median-filtered version of the image. Tests on datasets with known disks showed that this approach is more effective than ADI at recovering azimuthally symmetric disks \citep{Canovas2017}, which are strongly affected by ADI self-subtraction, and performs better at larger angular separations \citep[see also][]{2018MNRAS.480..236Z}. The inner region (within $\sim1$–$4$ FWHM), however, remains dominated by speckle noise and imperfect stellar PSF subtraction.

\begin{figure}
\centering

\includegraphics[width=0.9\columnwidth,trim=0 2.1cm 0 2.5cm,clip]{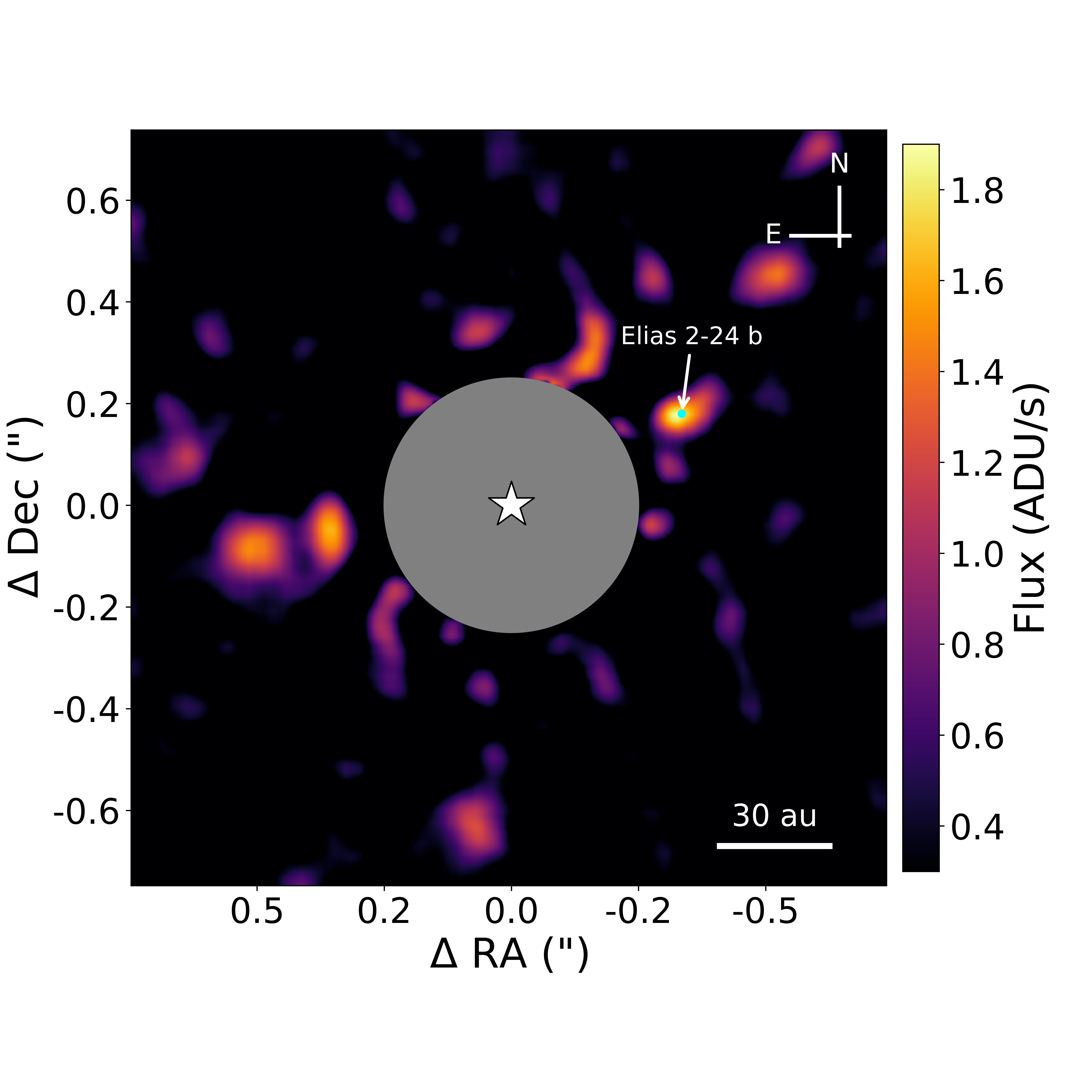}
\includegraphics[width=0.9\columnwidth,trim=0 2.1cm 0 2.5cm,clip]{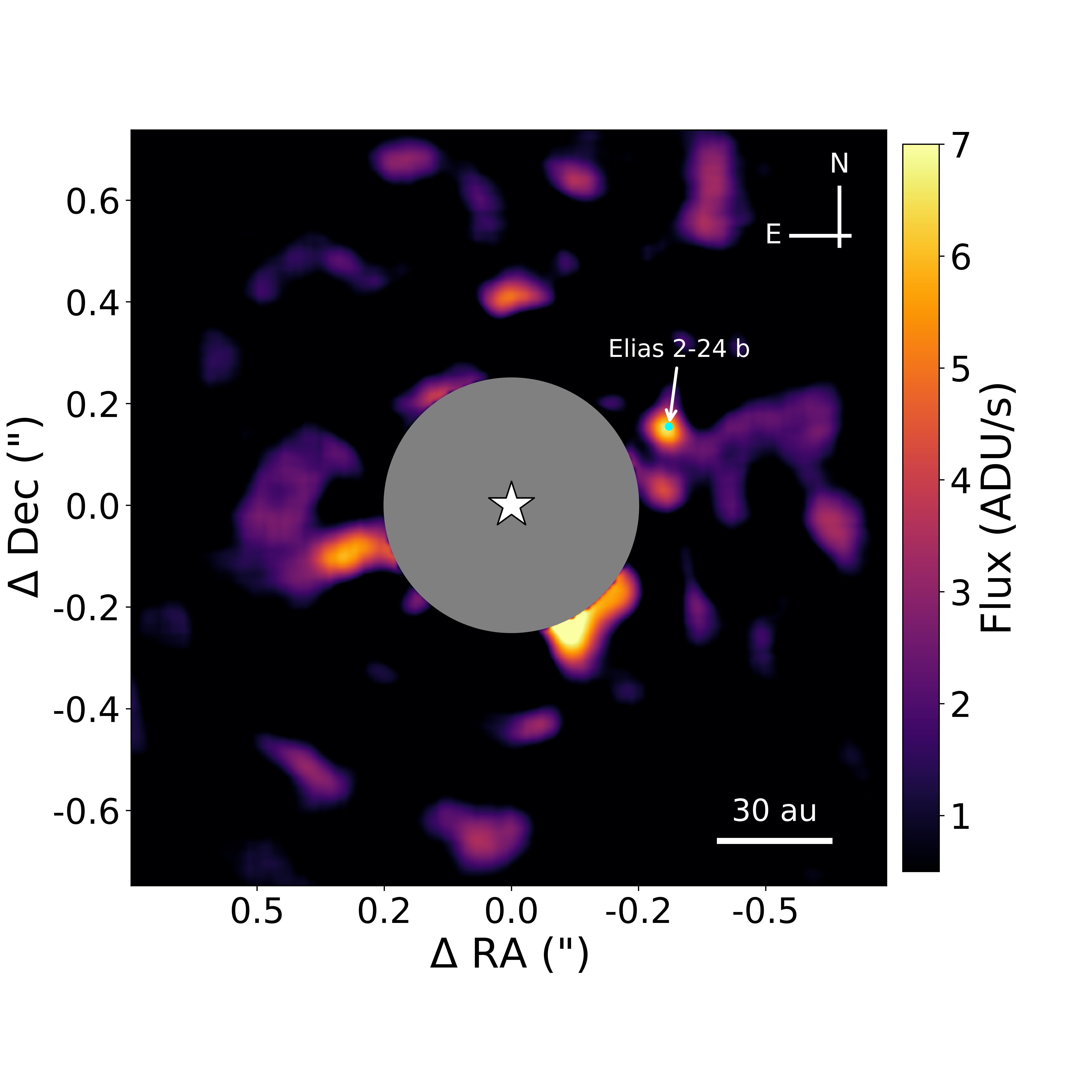}
\includegraphics[width=0.9\columnwidth,trim=0 2.1cm 0 2.5cm,clip]{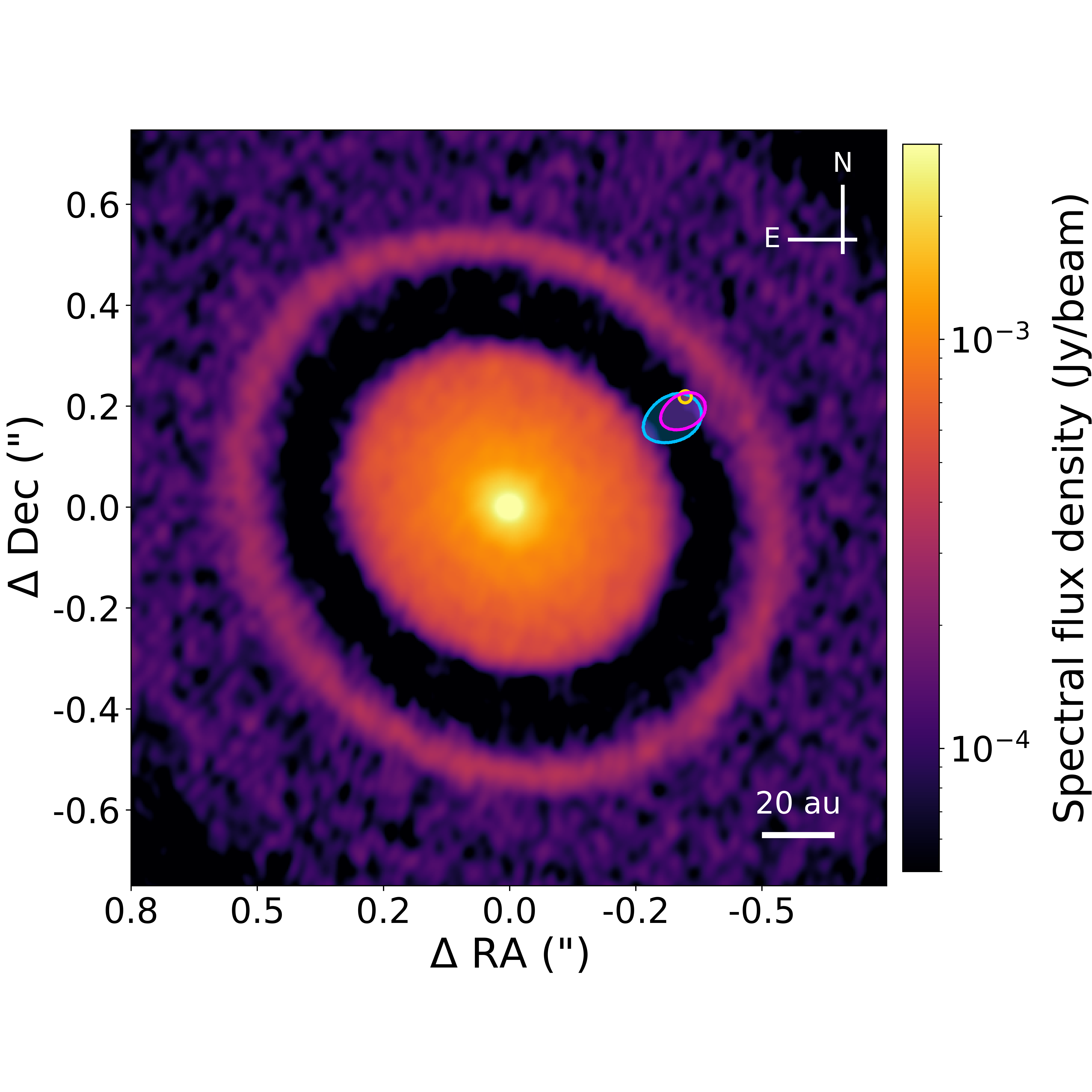}

\caption{Upper panel: annular ADI-PCA image of Elias 2-24 in the $L'$ band for the 21 June 2018 epoch. Middle panel: annular ADI-PCA image of Elias 2-24 for the 1 June 2020 observation. In both images the position of Elias 2-24 b is indicated by a white arrow and the inner $0.25^{\prime\prime}$ region is masked due to residual speckle noise. Lower panel: ALMA 1.3 mm continuum observation from the DSHARP survey \citep{Andrews2018}. The magenta, blue, and gold regions represent the area enclosing the position of the companion according to our results, and those of \citet{Pinte2023b} and \citet{Jorquera2021}, respectively. Elias 2-24 b is responsible for the gap and the accumulation of dust at the inner edge of the outer disk.
}
\label{elias224}

\end{figure}

\section{Confirmation of Elias 2-24 b}

\label{sec:results}
In this section we present the results of our analysis. We derived ADI-PCA post-processed images for all targets and subsequently investigated them for possible point-source detections and/or disk emission, also deriving 5$\sigma$ detection limits for the whole sample. We report no clear disk emission detections using either the ADI or RDI techniques, with only Elias 2-24 and PDS 70 (more details in Appendix \ref{app:pds70}) exhibiting point-like sources in our reductions.

\subsection{Detection limits}

We computed the $5\sigma$ contrast curves for each target (Figure \ref{contrasts}, left, and Table \ref{contrast_table} in Appendix \ref{app:contrasts}) using the \texttt{AppleFy} package \citep{Bonse2023}, based on the ADI-PCA procedure implemented in \texttt{VIP}. This approach accounts for small-sample statistics \citep{Mawet2014} by adopting a t-distribution for the test statistic under the assumption of Gaussian noise. A key advantage of \texttt{AppleFy} is the ability to evaluate multiple numbers of principal components simultaneously and derive an optimal curve by selecting the deepest contrast at each separation. The corresponding mass detection limits are shown in Figure \ref{contrasts} (right), obtained by converting the contrast curves using the ages and distances listed in Table \ref{obs}, together with the ATMO-NEQ-STRONG 2023 evolutionary models \citep{Chabrier2023}. These models are well suited to describe out-of-equilibrium atmospheres, which are expected in our targets given the extreme youth of the sample. Our results allow us to exclude the presence of additional companions beyond those already known. For PDS 70, we rule out planets with masses of $\sim7$–$10$ M\textsubscript{Jup} out to separations of $\sim300$ au. For the other targets, we exclude companions with masses between $\sim1.4$ and $8.1$ M\textsubscript{Jup} out to $\sim400$ au, depending on the system distance. For HD 98800 B, the contrast curve starts at $\sim0.5^{\prime\prime}$ ($\sim22.5$ au) because the innermost region is dominated by the stellar binary. The deepest contrast is achieved for Elias~2-24, where we reach $5\sigma$ detection limits of $\sim1.4$–$4.9$ M\textsubscript{Jup} between $\sim12$ and $370$ au. For this target, no uncertainty on the detection limits is reported, since we adopted an upper-limit age of 1 Myr. A more detailed discussion of this system's age is presented in Section \ref{sec:characterization}.

\subsection{Elias 2-24}
\label{subsec:elias}

We recovered the source presented by \citet{Pinte2023b} and \citet{Jorquera2021} in our Keck observation (Figure \ref{elias224}, up), and estimated its position as described in Section \ref{sec:obs_red}. We measured a separation of $394\pm31$ mas (corresponding to $54.9\pm4.3$ au) and a position angle of $298.8\pm3.2^\circ$. The measured position is consistent with previous studies. \citet{Jorquera2021} reported a separation of $0.411\pm0.008^{\prime\prime}$ and a position angle of $302.1\pm1.1^\circ$ (possibly the uncertainties presented for the astrometric position included only the errors on the fit and are likely underestimated), while \citet{Pinte2023b} derived a separation of $0.37\pm0.04^{\prime\prime}$ and a position angle of $298.4\pm4.6^\circ$. Our Keck astrometry agrees within $1\sigma$ with previous studies, consistently placing the candidate within a gap of the circumstellar disk observed in 1.3 mm continuum emission. As shown in Figure \ref{elias224} (bottom), the positions from the three datasets overlap, confirming their consistency.

We also re-reduced and analyzed the 1 June 2020 NIRC2 observation of Elias 2-24 (program C249, PI D. Mawet) presented by \citet{Wallack2024}. We re-detected the source at a separation of $351\pm30$ mas and a position angle of $294.3\pm4.8^\circ$. The astrometry derived from both epochs is reported in Table~\ref{elias_astro}, while the ADI-PCA post-processed image of the 2020 epoch is shown in the middle panel of Figure \ref{elias224}.

\begin{table}
\centering
\caption{Astrometry of Elias 2-24 b.}
\begin{tabular}{ccc}
\hline\hline
  Epoch  &	Separation     &    Position angle  \\
    (UT)    &  (mas)        &   ($^\circ$)   \\  
\hline
$2018.468$ &  $394\pm31$  & $298.8\pm3.2$ \\[0.75ex]
$2020.415$ &  $351\pm30$  & $294.3\pm4.8$ \\[0.75ex]
\hline
\end{tabular}
\label{elias_astro}
\end{table}

To assess the robustness of our detection in both epochs, we performed several tests. First, we applied different PCA approaches (as described in Section \ref{sec:obs_red}) over a wide range of subtracted PCs, and estimated the detection significance in each case. The source is consistently recovered in both annular and full-frame PCA reductions of the 2018 epoch. In the annular PCA case, the significance exceeds $3\sigma$ for more than 6 subtracted PCs and reaches a maximum of $3.85\sigma$ (SNR of 4.49) for 27 PCs, while the full-frame PCA reduction yield a peak significance of $3.69\sigma$ (SNR of 4.28) for 6 PCs. The companion was recovered with a lower significance, peaking at $2.82\sigma$ (SNR of 3.11) for 5 PCs, in the 2020 epoch. The lower significance of the detection, despite the longer integration time and larger field rotation compared to the 2018 epoch, is most likely due to the poorer atmospheric conditions during the 2020 observation. The seeing on 1 June 2020 was indeed $0.84^{\prime\prime}$, compared to $0.5^{\prime\prime}$ on 21 June 2018 (Table \ref{obs}). This likely reduced the sensitivity of the observation, an interpretation further supported by the slightly poorer contrast achieved at $\sim0.4^{\prime\prime}$, close to the candidate's separation \citep{Wallack2024} and at larger separations.

We then removed the companion from the reduced data cube by injecting a negative companion with the parameters derived from the MCMC analysis, and re-processed the data using the different PCA approaches. In all cases, the post-processed images showed residual noise at the location of the companion consistent with that observed at the same radial separation, with no significant residual structures remaining. Finally, we injected a synthetic positive companion with the MCMC-derived parameters at a different position angle and verified whether it could be recovered, finding that the injected companion is robustly recovered. Although the detections do not reach the standard $5\sigma$ confidence threshold, the combination of these results with the previous tests strengthens the interpretation of the observed signals as genuine detections rather than residual speckle noise.

\begin{figure}[ht!]
 \centering
    \includegraphics[width=0.7\columnwidth]{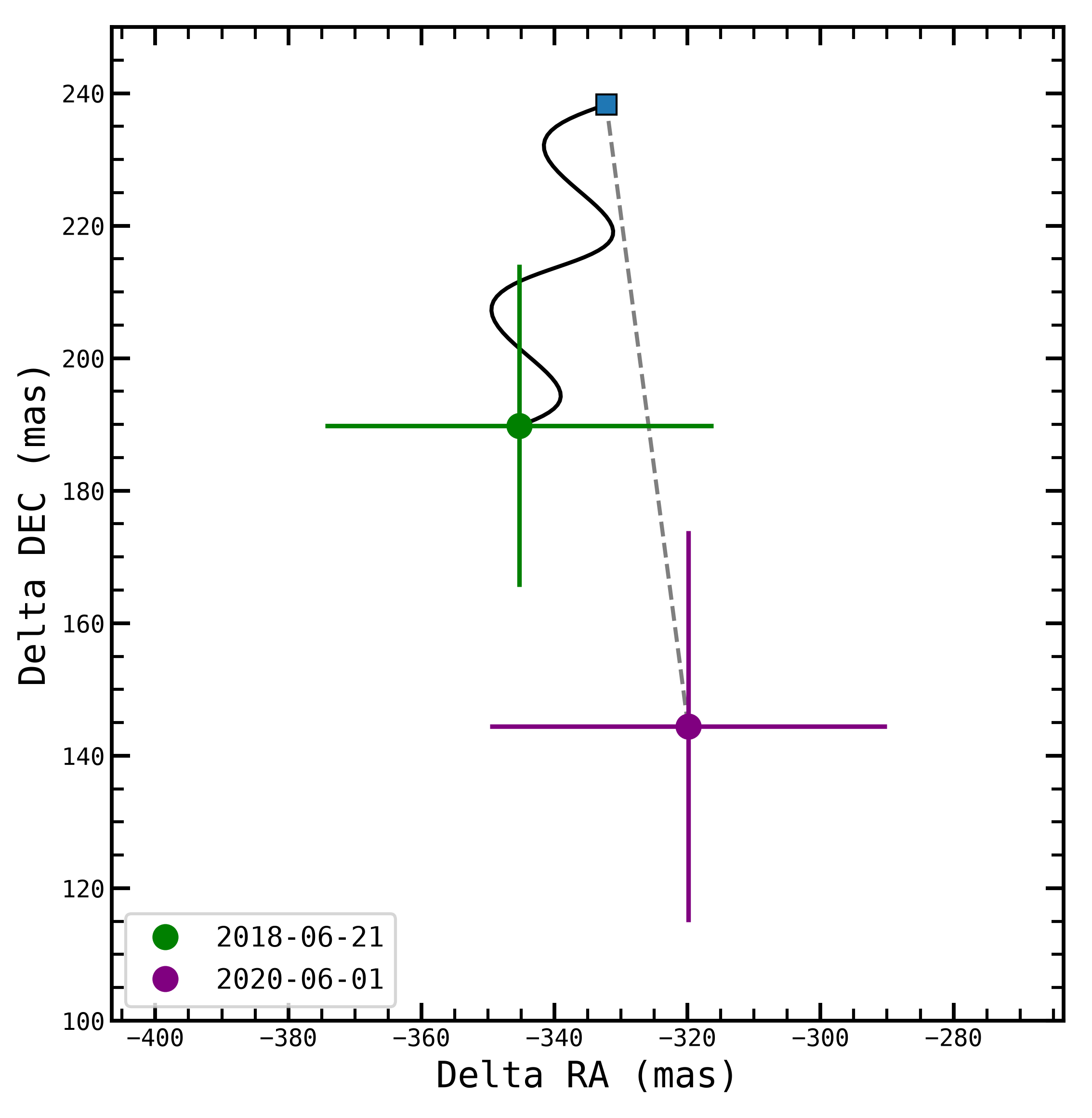}
    \caption{Proper motion analysis of Elias 2-24 b. The green circle represents the companion's measured position in the 2018 epoch and the purple circle represents the companion's measured astrometry in the 2020 epoch. The solid black line shows the expected motion of a static background source relative to the host star, while the blue square represents the expected position for a static background source at the epoch of the second observation.}
    \label{propermotiontest}
\end{figure}

To verify whether the candidate's astrometry is compatible with that of a bound companion, we first compare its position in the 2020 epoch with the expected position of a static background object (Figure \ref{propermotiontest}). The comparison between the expected position of a background source, shown as a blue square, and the position of the candidate, shown as a purple circle, shows that the candidate is incompatible with the static background hypothesis at the 3.2$\sigma$ confidence level.

We then quantify the probability -- expressed in terms of a false alarm probability (FAP) -- that a background source could exhibit an astrometric displacement consistent with that expected for a bound companion. We first generated a synthetic sample of background stars using the Besançon Galaxy Model interface\footnote{\url{https://model.obs-besancon.fr/modele_home.php}} \citep{Czekaj2014}, selecting stars at distances between 0 and 50 kpc within a 5 arcmin radius of the position of Elias 2-24. We further selected stars with apparent $L'$ magnitudes comparable to that of the candidate (15–17 mag), assuming an interstellar extinction of $A_V=8.7$ mag (more details in Section \ref{sec:characterization}).

To compute the FAP, we define an "interesting region" in proper-motion space, within which a background star could show a motion compatible with that expected for a bound companion (e.g., \citealt{Squicciarini2022}). Based on our observations, we consider all background sources with negative proper motion in declination as "interesting", while imposing no constraints on their proper motion in right ascension. Out of 668 synthetic stars, only two fall within this region, yielding a FAP of $2.99\cdot10^{-3}$, corresponding to a $2.8\sigma$ confidence level. Finally, we compute the FAP of finding such a background source within our field of view. Considering a field of view of $2.6^{\prime\prime}$, we derive a FAP of $1.50\cdot10^{-4}$, corresponding to a $3.6\sigma$ significance. We therefore conclude that the candidate is physically bound to Elias 2-24 at a high level of confidence.

The consistency among the multiple observations, together with the repeated detection of the source at a compatible location across different instruments and observatories and the common proper motion analysis, allows us to confirm the planetary nature of the candidate, hereafter referred to as Elias 2-24 b.

Finally, we estimated the angular motion of Elias 2-24 b between our 2018 Keck epoch and 1 June 2026, assuming a circular orbit coplanar with the disk. Adopting a disk inclination of $23.6^\circ$ and a position angle of $43.4^\circ$ \citep{Cieza2017}, together with a stellar mass of $0.78^{+0.35}_{-0.13}$ M\textsubscript{$\odot$} \citep{Andrews2018}, we infer a predicted motion of $6.2^{+2.2}_{-1.1}$$^\circ$. This displacement is well above our $1\sigma$ confidence interval and should be clearly resolvable. Current and future observations will therefore be able to probe the Keplerian motion of the planet, providing a first, less assumption-dependent estimate of its orbital parameters. As of now, we are not yet able to probe the Keplerian motion of Elias 2-24 b, despite its re-detection in the 2020 Keck epoch. Assuming a circular orbit, the expected angular motion between the two epochs is only $\sim1.5^\circ$, which is smaller than our uncertainties.

\section{Characterization of Elias 2-24 b}
\label{sec:characterization}

We derive an $L'$ contrast for Elias 2-24 b of $9.86^{+0.53}_{-0.87}$ mag by estimating the flux ratio between the companion and the host star as described in Section~\ref{sec:obs_red}. \citet{Jorquera2021} measured an $L'$-band contrast of $8.81\pm0.12$ mag, while \citet{Pinte2023b} obtained a value of $8.78^{+0.98}_{-0.48}$ mag. Our estimate is slightly fainter, although still consistent within $1\sigma$ with \citet{Pinte2023b} and within $1.2\sigma$ with \citet{Jorquera2021}.

In order to estimate the absolute magnitude of the planet, we combine our derived $L'$ contrast and the distance of $139.3$ pc reported in Table~\ref{obs}, while accounting for interstellar extinction. As reported in \citet{Cieza2021}, Elias 2-24 has an extinction of $A_V=8.7$ mag, corresponding to $0.5$ mag in the $L'$ band. Assuming that the star and the planet experience the same extinction, we derive an absolute $L'$ magnitude of $10.08^{+0.53}_{-0.87}$ mag for Elias 2-24 b.

We then estimated the evolutionary mass and effective temperature of Elias 2-24 b by assuming an age of 1 Myr, while noting that the system may be even younger. \citet{Andrews2010} reports an age of 0.4 Myr, while \citet{Andrews2018} adopts an age of $0.20^{+0.30}_{-0.12}$ Myr, the youngest for the entire DSHARP sample. Elias 2-24 also has the highest accretion rate reported in DSHARP ($\log \dot{M}_\star=6.4\pm0.4\,M_\odot$yr$^{-1}$; \citealt{Andrews2018}) and the brightest disk in the ODISEA survey (361 mJy at 230 GHz; \citealt{Williams2019}). Furthermore, \citet{Evans2009} report a bolometric temperature of 980 K, indicating extreme youth for a T Tauri star. Overall, even though deriving reliable ages for young stellar objects younger than 1 Myr is notoriously difficult, we conclude that the Elias 2-24 system is extremely young and adopt an age of $\lesssim1$ Myr in our work. This short timescale is consistent with the fact that some gaps in disks appear to emerge in the embedded phase of young stellar objects (age $\lesssim$ 1 Myr) and become ubiquitous by the SED Class II stage (age $\gtrsim$ 1 Myr) in bright and massive disks, as found by high-resolution ALMA surveys (\citealt{Sheehan2020,Shoshi2025,Hsieh2025,Bosschaart2026}).

To investigate the dependence of the evolutionary mass and effective temperature (T\textsubscript{eff}) of Elias 2-24 b on the initial conditions (e.g., the initial entropy) and atmospheric properties, we estimated these quantities using different evolutionary models. First, we adopted the ATMO-NEQ-STRONG 2023 models \citep{Chabrier2023}, which assume a non-equilibrium atmosphere with strong vertical mixing and a high initial luminosity ("hot-start" scenario), deriving a mass of $2.54^{+1.66}_{-0.66}$ M\textsubscript{Jup} and an effective temperature of $1315^{+357}_{-165}$ K. We then considered the AMES evolutionary models \citep{Baraffe2003}, both assuming a hot-start formation scenario but differing in their treatment of atmospheric clouds: a cloud-free atmosphere (AMES-COND; \citealt{Baraffe1998}) and a cloudy atmosphere (AMES-DUSTY; \citealt{Chabrier2000}). With the AMES-COND models we infer a mass of $3.84^{+0.99}_{-2.03}$ M\textsubscript{Jup} and an effective temperature of $1517^{+249}_{-325}$ K while, for AMES-DUSTY, since the central and fainter magnitudes of the planet lie outside the model grid, we are only able to put upper limits of 3.97 M\textsubscript{Jup} for the mass and of 1608 K for the effective temperature. Finally, to explore the dependence on the assumed initial entropy, we used the BEX models \citep{Mordasini2017,Marleau2019,Linder2019}, which employ the \texttt{completo21} cooling evolution code with boundary conditions provided by the AMES-COND/DUSTY atmospheric models. We considered the "coldest-start", "warm-start", "hot-start", and "hottest-start" scenarios, corresponding to different assumptions for the initial luminosity of the planet. At the age and luminosity of Elias 2-24 b, the coldest-start and warm-start model grids converge to the same solution. For the coldest-start/warm-start scenarios, we derive masses of $13.84^{+1.91}_{-1.08}$ M\textsubscript{Jup} and $13.13^{+2.43}_{-1.35}$ M\textsubscript{Jup}, with effective temperatures of $1667^{+339}_{-200}$ K and $1536^{+434}_{-244}$ K, for the COND and DUSTY models, respectively. For the hot-start scenario, the inferred masses decrease to $4.89^{+3.97}_{-1.50}$ M\textsubscript{Jup} and $3.81^{+4.44}_{-1.41}$ M\textsubscript{Jup}, with effective temperatures of $1505^{+378}_{-186}$ K and $1359^{+472}_{-212}$ K, for the COND and DUSTY models, respectively. Finally, for the hottest-start scenario, we obtain masses of $3.70^{+2.58}_{-1.07}$  M\textsubscript{Jup} and $2.93^{+2.91}_{-1.03}$ M\textsubscript{Jup}, with effective temperatures of $1474^{+345}_{-184}$ K and $1331^{+436}_{-202}$ K, for the COND and DUSTY models, respectively. All the results are summarized in Table \ref{masses}.

\begin{table}
\centering
\caption{Evolutionary mass and effective temperature estimates for Elias 2-24 b obtained with different models.}
\begin{tabular}{lcc}
\hline\hline
  Model & Mass  &	T\textsubscript{eff}  \\
    &    (M\textsubscript{Jup})    &  (K)    \\  
\hline
ATMO-NEQ-STRONG 2023 &  $2.54^{+1.66}_{-0.66}$  & $1315^{+357}_{-165}$ \\[0.75ex]
AMES-COND &  $3.84^{+0.99}_{-2.03}$  & $1517^{+249}_{-325}$ \\[0.75ex]
AMES-DUSTY &  $\lesssim3.97$  & $\lesssim1608$ \\[0.75ex]
BEX-COND (coldest/warm start) &  $13.84^{+1.91}_{-1.08}$  & $1667^{+339}_{-200}$ \\[0.75ex]
BEX-COND (hot start) &  $4.89^{+3.97}_{-1.50}$  & $1505^{+378}_{-186}$ \\[0.75ex]
BEX-COND (hottest start) &  $3.70^{+2.58}_{-1.07}$  & $1474^{+345}_{-184}$ \\[0.75ex]
BEX-DUSTY (coldest/warm start) &  $13.13^{+2.43}_{-1.35}$  & $1536^{+434}_{-244}$ \\[0.75ex]
BEX-DUSTY (hot start) &  $3.81^{+4.44}_{-1.41}$  & $1359^{+472}_{-212}$ \\[0.75ex]
BEX-DUSTY (hottest start) &  $2.93^{+2.91}_{-1.03}$  & $1331^{+436}_{-202}$ \\[0.75ex]

\hline
\end{tabular}
\label{masses}
\tablecomments{These estimates do not account for the effects of accretion and should therefore be considered as upper limits.}
\end{table}

Before discussing these results, it is important to highlight a fundamental point. All these models neglect accretion effects, assuming that the observed luminosity is entirely due to the internal luminosity of the planet. However, as pointed out by several studies (e.g., \citealt{Mordasini2017}), at such young ages the accretion luminosity can dominate the intrinsic luminosity of the planet. This results in an overestimation of the internal luminosity and, consequently, of the inferred evolutionary mass and effective temperature. Therefore, the values reported in Table \ref{masses} should be interpreted as upper limits rather than as definitive estimates.

Beyond the different atmospheric assumptions, the choice of evolutionary model has deeper implications in this specific case. As shown in Table \ref{masses}, the inferred effective temperature of Elias 2-24 b ranges from approximately 1100 to 2000 K, spanning the classical temperature regime in which the L–T transition occurs (e.g., \citealt{Vos2019}). This transition is generally attributed to the settling of clouds in the atmosphere, which is characteristic of L-type objects, while cooler T-type objects are expected to have less cloudy atmospheres. These two extreme cases, represented by cloudy and dust-free atmospheres, are captured by the AMES-DUSTY model (closer to the L sequence) and the AMES-COND model (closer to the T sequence), respectively. However, the temperature at which the L–T transition occurs depends on several properties, such as surface gravity. Lower-gravity objects undergo the L–T transition at lower temperatures (down to $\sim$800-1000 K; \citealt{Gratton2024}) than higher-gravity objects (e.g., \citealt{Faherty2016}), and they appear significantly redder, likely due to stronger vertical mixing (e.g., \citealt{Miles2023}). Consequently, young substellar objects with L spectral types appear redder than their older counterparts \citep{Liu2016}. Because of its extreme youth, the properties of Elias 2-24 b are likely better described by dusty and non-equilibrium models, although the L–T transition in these regimes is not yet well reproduced by current evolutionary models. These models yield lower masses and temperatures compared to cloud-free models (Table \ref{masses}). However, additional observations, possibly at different wavelengths, and spectroscopic data are needed to better constrain its properties (such as its temperature and surface gravity) and enable a spectral characterization of this planet.

\begin{figure*}[t!]
 \centering
 \includegraphics[width=0.49\linewidth,trim=0 0 0 0cm,clip]{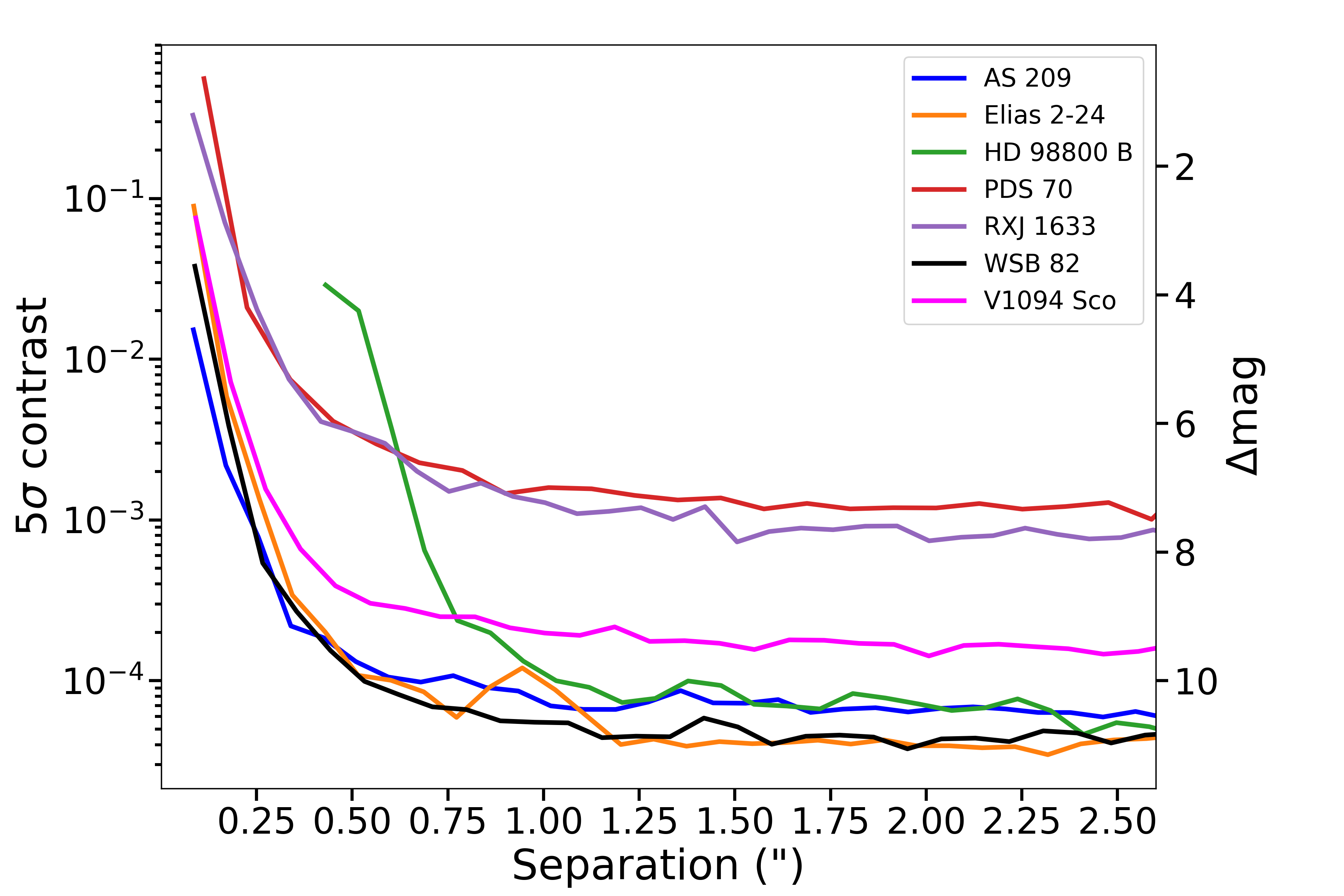}
 \includegraphics[width=0.49\linewidth,trim=0 0 0 0cm,clip]{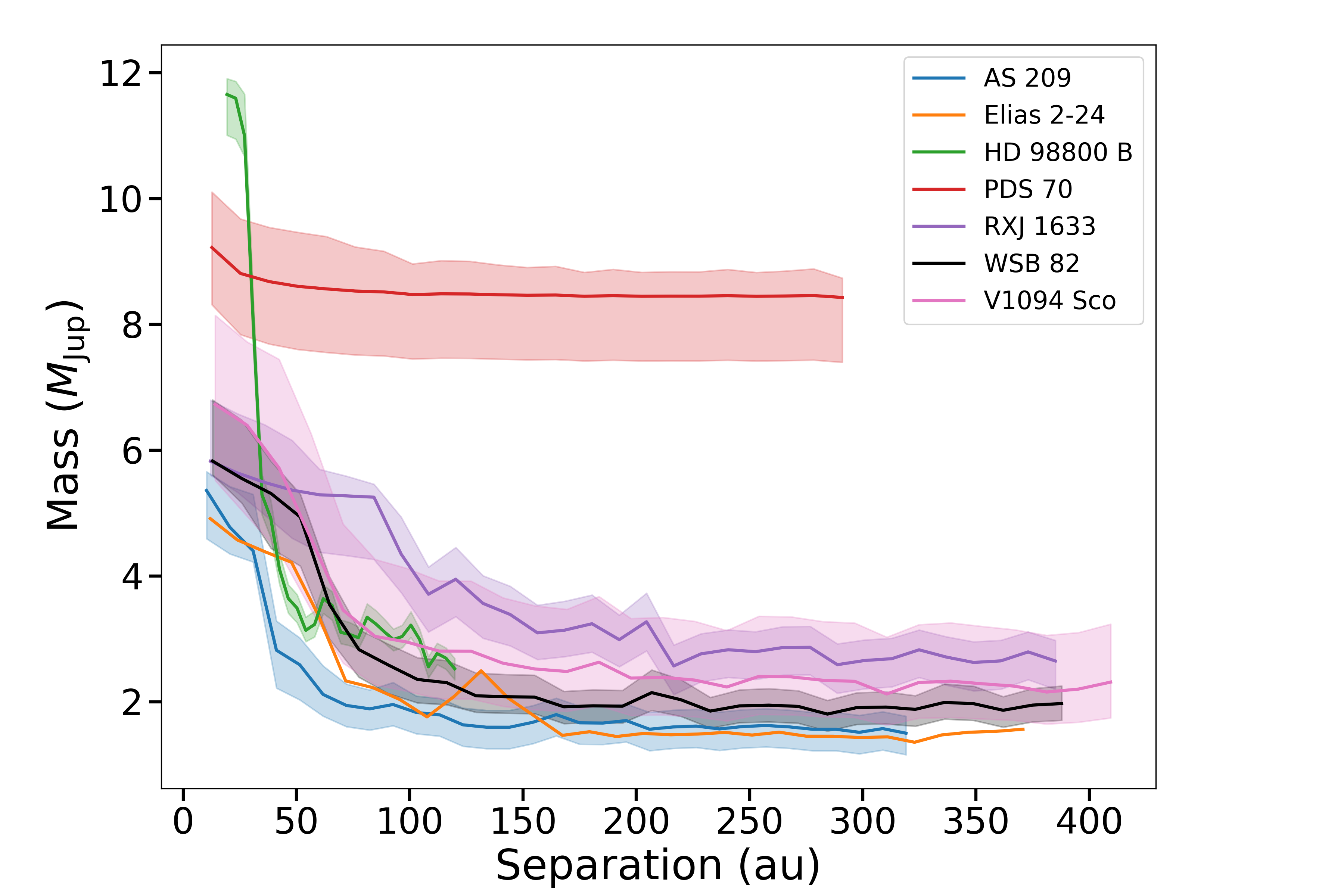}
    \caption{Left: 5$\sigma$ contrast curves in separation versus contrast and magnitude difference with respect to the host star. Right: corresponding detection limits converted into mass using the ATMO-NEQ-STRONG 2023 models. The uncertainty on the mass limits reflects the uncertainty on the host star's age.}
    \label{contrasts}
\end{figure*}

\section{Discussion}
\label{sec:discussion}

The gap of Elias 2-24, centered at $\sim0.4^{\prime\prime}$ (55 au) and with a width of $\sim30$ au, was first reported by \citet{Cieza2017} using ODISEA observations with a modest angular resolution of 0.2$^{\prime\prime}$. Based on numerical arguments relating the gap width to the Hill radius of the putative planet, the authors estimated a planet mass in the range of 1–8 M\textsubscript{Jup}. Using hydrodynamical models, \citet{Dipierro2018} and \citet{Zhang2018} later refined this estimate toward the lower end of the range, around $\sim1$ M\textsubscript{Jup}. In the latter case, the estimate was derived using higher-resolution ($\sim0.05^{\prime\prime}$) DSHARP observations \citep{Andrews2018}. 

The direct detection of a planet candidate within the gap of Elias 2-24 was first reported by \cite{Jorquera2021} in the near-IR ($L'$-band) using VLT/NACO data. Deep ALMA $^{12}$CO observations of Elias 2-24 later revealed a bright spot close to the location of the point source, within the continuum gap \citep{Pinte2023b}. This last study also pointed out that the morphology of the bright spot resembles that predicted by \citet{Perez2015} for circumplanetary disk emission, further supporting the planetary interpretation of the detection. We recover the candidate in two different Keck/NIRC2 observations and confirm its planetary nature in Section \ref{subsec:elias}.

Using the \citet{Baraffe2003} isochrones, \citet{Jorquera2021} estimated a mass of $\sim5$ M\textsubscript{Jup}, reduced to $0.4$–$1.72$ M\textsubscript{Jup} when accounting for accretion effects with the \citet{Zhang2018} models. Re-analyzing the same observations, and comparing the observed flux with $^{12}$CO kinematics, \citet{Pinte2023b} concluded that the data are consistent either with a non-accreting $\sim8$ M\textsubscript{Jup} planet or with an accreting $\sim3$ M\textsubscript{Jup} planet. In reality, Elias 2-24 b is likely undergoing vigorous accretion, favoring estimates in the lower mass range. In the five-stage evolutionary sequence of disk substructures proposed by \citet{Cieza2021}, the substructures seen at high-resolution ALMA continuum observations can be understood in terms of the effect that (proto)planets forming through core accretion have on the evolution of the disk. This interpretation is supported by numerical models of disk evolution and planet disk interactions \citep{Orcajo2025}. These models show that planetary embryos of just $0.1\, M_\oplus$ can produce inflection points in the observed brightness profiles at mm wavelengths but that fully-grown giant planets are needed to carve detectable gaps, which are originally narrow and eventually become cavities when the inner disks dissipate. These models use Elias 2-24 as the archetype “Stage III”, corresponding to the envelope accretion phase, when protoplanets are expected to undergo an accretion-driven luminosity spike \citep{Marley2007}. As the planet grows, it produces a strong pressure bump, which traps mm particles at the outer edge of the gap,  resulting in the characteristic bright rim that is seen in the ALMA image shown in Figure \ref{elias224} (lower panel). According to the models, the relatively narrow gap in Elias 2-24 can be produced by a 1 M\textsubscript{Jup} planet, and  the “Stage III” configuration lasts of the order of 0.1 Myr before the inner disk dissipates and the system transitions onto a “Stage IV” structural evolution \citep{Orcajo2025}.

We also provide a new estimate of the planet mass accretion rate. Comparing our estimate for the absolute magnitude of Elias 2-24 b (in Section \ref{sec:characterization}) with the models of \citet{Zhu2015}, and assuming an inner circumplanetary disk radius of 2 R\textsubscript{Jup}, we derive $M_p\dot{M}=2.78\cdot10^{-6}$ M\textsubscript{Jup}$^2$ yr$^{-1}$ through logarithmic interpolation. Adopting the mass predictions from the hydrodynamical models of \citet{Zhang2018}, we then estimate accretion rates for two masses: 0.41 and 1.72 M\textsubscript{Jup}. The resulting accretion rates are $6.77\cdot10^{-6}$ and $1.61\cdot10^{-6}$ M\textsubscript{Jup} yr$^{-1}$, respectively. As noted by \citet{Pinte2023b}, strong cloud contamination makes a precise mass determination difficult. Improved constraints could be obtained with higher-precision observations, for example with VLT/ERIS. 

Independently of the exact mass of Elias 2-24 b, its confirmation has strong implications for planet formation because it demonstrates that a Jovian planet can be formed by core accretion at  $\sim$50 au in just $\lesssim$1 Myr, much faster than what can be inferred from the PDS 70 and WISPIT 2 systems, which are both $\sim$5 Myr old \citep{Muller2018, Close2025}. This makes Elias 2-24 b the youngest planet detected through high-contrast imaging so far. The close agreement of the disk morphology of Elias 2-24 with the core accretion model, and in more general terms, the demographic of disk substructures, showing that gaps and cavities \citep{Andrews2020,GuerraAlvarado2025,Bhowmik2026} are overwhelmingly more common than the fragmenting spiral arms associated with gravitational instability \citep{Zhu2012,Tobin2016,Weber2025} strongly suggests that Elias 2-24 b and the other protoplanets directly imaged in the gaps and cavities of other nearby disks (e.g., PDS 70 and WISPIT 2) are most likely formed via core accretion. Strictly speaking, it is possible that a giant planet could form by gravitational instability in a fragmenting disk, very early on and at large distances,  and then migrate inward (together with the gap) to a region of the disk showing no evidence of fragmentation. However, this scenario seems unlikely since there is no evidence of gap migration either. On the contrary, the disk sizes measured at mm continuum are believed to trace the location of the pressure bumps in a given disk \citep{Rosotti2019} and the results by \citet{Dasgupta2025b} show disk sizes in the Ophiuchus molecular cloud do not seem to evolve with time (the distribution of continuum  disk sizes peaks at 14 au for both very young embedded disks and older, more evolved, Class II sources). If the pressure bumps halting dust migration are produced by (proto)planets as the models and the observations seem to indicate,  then these results place strong constraints on the (lack of) migration for both (proto)planets and their accompanying gaps.

Despite the likely core-accretion formation of Elias 2-24 b, a mechanism traditionally associated with cold-start models, its properties are better reproduced by hot-start models, because the planet is still accreting (e.g., \citealt{Marley2007}). Indeed, hydrodynamical models by \citet{Zhang2018}, based on the observed properties of the disk gap (e.g., its width, edge ellipticity, and asymmetry), infer a planetary mass between 0.41 and 1.72 M\textsubscript{Jup} to reproduce the observed disk structure. This range rules out the upper mass estimates obtained with the coldest-start and warm-start models (Table~\ref{masses}), while remaining compatible, at least within the lower end of the uncertainties, with the values inferred from the ATMO-NEQ-STRONG 2023 and the BEX-DUSTY hot-start and hottest-start models. More massive planets would produce disk structures inconsistent with the observations and can therefore be excluded. This apparent contradiction is, however, consistent with more recent studies, which show that a core-accretion origin cannot be ruled out based on a high observed luminosity alone. Indeed, if the planetary core is sufficiently massive \citep{Mordasini2013}, if the accretion shock is relatively inefficient at cooling the infalling gas \citep{Mordasini2017}, or if the shock dissipates sufficient heat into the accreting gas \citep{Marleau2017,Marleau2019b}, a young giant planet formed through core accretion may still exhibit a hot-start-like luminosity (e.g., \citealt{Nowak2020}). We therefore conclude that the upper mass and effective temperature limits of Elias 2-24 b are better described by hot-start, dusty evolutionary models. Based on its observed luminosity, we infer an upper mass limit in the range of 1.9-4.0 M\textsubscript{Jup} (Table \ref{masses}), ruling out the higher-mass solutions.

Our inferred evolutionary mass is lower than previous non-accreting estimates, differing by $2.0\sigma$ from \citet{Jorquera2021} and $5.0\sigma$ from \citet{Pinte2023b}. We also use the \citet{Zhu2015} models to estimate the accretion rate for a mass of 3.0 M\textsubscript{Jup}, corresponding to our central evolutionary mass estimate, deriving a value of $\dot{M}=9.26\cdot10^{-7}$ M\textsubscript{Jup} yr$^{-1}$.

\section{Conclusion}
\label{sec:conclusion}

In this work, we present Keck/NIRC2 vortex coronagraph observations of seven stars hosting protoplanetary disks with known substructures potentially linked to forming protoplanets. Our goal is to investigate whether the companions responsible for these features can be directly imaged. All targets were observed between May and June 2018 in the $L'$ band, except for PDS 70, which was observed in the $M_s$ band. Finally, we derived $5\sigma$ contrast curves, which allow us to place upper limits on the masses of potential planets within these disks. By converting the $5\sigma$ contrasts into masses using the ATMO-NEQ-STRONG 2023 models \citep{Chabrier2023}, we reach sensitivities down to $\sim$1–2 $M\textsubscript{Jup}$ for several systems in our sample.

We re-detect Elias 2-24 b, the planet candidate previously reported by \citet{Jorquera2021} and \citet{Pinte2023b}, in our Keck observations, confirming its planetary nature through an independent dataset and reduction. We measure a separation of $394\pm31$ mas and a position angle of $298.8\pm3.2^\circ$, consistent with previous studies. We also re-detect the planet in a subsequent 2020 Keck observation at a separation of $351\pm30$ mas and a position angle of $294.3\pm4.8^\circ$, albeit at a lower significance.

Comparing the derived $L'$ contrast with several evolutionary models, and concluding that the properties of Elias 2-24 b are better described by non-equilibrium, hot-start models, we infer a mass between 1.9 and 4.0 M\textsubscript{Jup}. However, this estimate should be treated with caution and regarded as an upper limit because of strong disk extinction and accretion effects. Future observations, for example with VLT/ERIS, could provide tighter mass constraints and probe the Keplerian motion of Elias 2-24 b, potentially enabling the first orbital constraints. Among the exoplanet population, Elias 2-24 b stands out for several reasons. It strengthens the connection, already suggested by the WISPIT 2 system, between narrow gaps in disks and protoplanets. The confirmation of Elias 2-24 b also supports the interpretation that Elias 2-24 represents a specific stage in the core accretion model, where planets are rapidly gaining mass by accreting their envelopes. This stage represents an epoch where forming planets are particularly bright and generate strong pressure bumps that have noticeable effects in the outer disks, where dust starts to accumulate in bright rims. In this context, Elias 2-24 b demonstrates that giant planets can form through core accretion at separations of $\sim50$ au within only $\sim1$ Myr, reinforcing the claims that planets are responsible for the gaps observed even at these young ages. Finally, it represents the youngest planet detected so far. Future observations of these systems with next-generation adaptive optics instruments and/or space-based facilities will be needed to directly detect smaller/closer putative companions. Coupled with more precise mass estimates independent of disk kinematics, such observations will be fundamental for achieving a deeper understanding of the observed disk substructures and their nature.

\begin{acknowledgments}
The authors acknowledge support from ANID -- Millennium Science Initiative Program -- Center Code NCN2024\_001. A.B. and A.Z. acknowledge support from Fondecyt Regular grant number 1250249. L.A.C. acknowledges support from ANID, FONDECYT Regular grant No. 1241056. S.P. acknowledges support from FONDECYT 1231663 and CIRAS-AI FIUF137139-USACH. This research has made use of the Keck Observatory Archive (KOA), which is operated by the W. M. Keck Observatory and the NASA Exoplanet Science Institute (NExScI), under contract with the National Aeronautics and Space Administration. This paper makes use of the following ALMA data: ADS/JAO.ALMA\#2013.1.00498.S. ALMA is a partnership of ESO (representing its member states), NSF (USA) and NINS (Japan), together with NRC (Canada), NSTC and ASIAA (Taiwan), and KASI (Republic of Korea), in cooperation with the Republic of Chile. The Joint ALMA Observatory is operated by ESO, AUI/NRAO and NAOJ. Part of this work was carried out at the Jet Propulsion Laboratory, California Institute of Technology, under a contract with the National Aeronautics and Space Administration (80NM0018D0004).
\end{acknowledgments}

\facilities{Keck/NIRC2, ALMA}

\software{\texttt{VIP} \citep{GomezGonzalez2017,Christiaens2023}, \texttt{AppleFy} \citep{Bonse2023}}

\appendix

\section{Table of contrast curves}
\label{app:contrasts}
Table \ref{contrast_table} shows optimal 5$\sigma$ contrast curves for each target, interpolated to several separations.  For HD 98800 B, our contrast curve begins at $\sim0.5^{\prime\prime}$, as the innermost region is dominated by the stellar binary.
\renewcommand{\arraystretch}{1.5}
\begin{sidewaystable*}
\centering
\caption{Optimal 5$\sigma$ contrast curves at selected separations.}
\resizebox{\textheight}{!}{
\begin{tabular}{lcccccccccccccccc}
\hline
\hline
Target & $0.1^{\prime\prime}$ & $0.2^{\prime\prime}$ & $0.3^{\prime\prime}$ & $0.4^{\prime\prime}$ & $0.5^{\prime\prime}$ & $0.6^{\prime\prime}$ & $0.7^{\prime\prime}$ & $0.8^{\prime\prime}$ & $0.9^{\prime\prime}$ & $1.0^{\prime\prime}$ & $1.25^{\prime\prime}$ & $1.5^{\prime\prime}$ & $1.75^{\prime\prime}$ & $2.0^{\prime\prime}$ & $2.25^{\prime\prime}$ & $2.5^{\prime\prime}$ \\ [0.75ex]
\hline \\
V1094 Sco & 6.03E-02 & 5.37E-03 & 1.21E-03 & 5.38E-04 & 3.45E-04 & 2.90E-04 & 2.60E-04 & 2.50E-04 & 2.18E-04 & 1.98E-04 & 1.87E-04 & 1.64E-04 & 1.77E-04 & 1.44E-04 & 1.65E-04 & 1.49E-04 \\ [0.75ex]
AS 209 & 1.08E-02 & 1.51E-03 & 3.96E-04 & 1.94E-04 & 1.37E-04 & 1.05E-04 & 1.00E-04 & 1.00E-04 & 8.80E-05 & 7.30E-05 & 7.15E-05 & 7.25E-05 & 6.53E-05 & 6.57E-05 & 6.50E-05 & 6.16E-05 \\ [0.75ex]
HD 98800 B & ... & ... & ... & ... & 2.15E-02 & 3.91E-03 & 5.68E-04 & 2.25E-04 & 1.65E-04 & 1.11E-04 & 7.53E-05 & 8.33E-05 & 7.16E-05 & 7.01E-05 & 7.53E-05 & 5.47E-05 \\ [0.75ex]
WSB 82 & 2.83E-02 & 2.33E-03 & 4.13E-04 & 2.02E-04 & 1.16E-04 & 8.58E-05 & 7.00E-05 & 6.59E-05 & 5.61E-05 & 5.50E-05 & 4.51E-05 & 5.21E-05 & 4.56E-05 & 4.07E-05 & 4.42E-05 & 4.18E-05 \\ [0.75ex]
Elias 2-24 & 5.72E-02 & 3.61E-03 & 6.82E-04 & 2.41E-04 & 1.20E-04 & 1.01E-04 & 8.06E-05 & 6.78E-05 & 1.04E-04 & 9.80E-05 & 4.18E-05 & 4.12E-05 & 4.17E-05 & 3.94E-05 & 3.79E-05 & 4.29E-05 \\ [0.75ex]
RX J1633.9-2442 & 2.44E-01 & 4.35E-02 & 1.14E-02 & 4.67E-03 & 3.56E-03 & 2.79E-03 & 1.80E-03 & 1.61E-03 & 1.46E-03 & 1.29E-03 & 1.19E-03 & 7.55E-04 & 8.70E-04 & 7.55E-04 & 8.79E-04 & 7.75E-04 \\ [0.75ex]
PDS 70 & 5.56E-01 & 4.35E-02 & 1.06E-02 & 5.37E-03 & 3.56E-03 & 2.72E-03 & 2.21E-03 & 1.96E-03 & 1.46E-03 & 1.57E-03 & 1.41E-03 & 1.30E-03 & 1.21E-03 & 1.19E-03 & 1.17E-03 & 1.22E-03 \\ [0.75ex] \\
\hline
\end{tabular}
}
\label{contrast_table}
\end{sidewaystable*}

\section{Additional results and discussion} 
\label{app:pds70}
\subsection{PDS 70}
\begin{figure}[ht!]
 \centering
    \includegraphics[width=\linewidth,trim=0 2cm 0 2.7cm,clip]{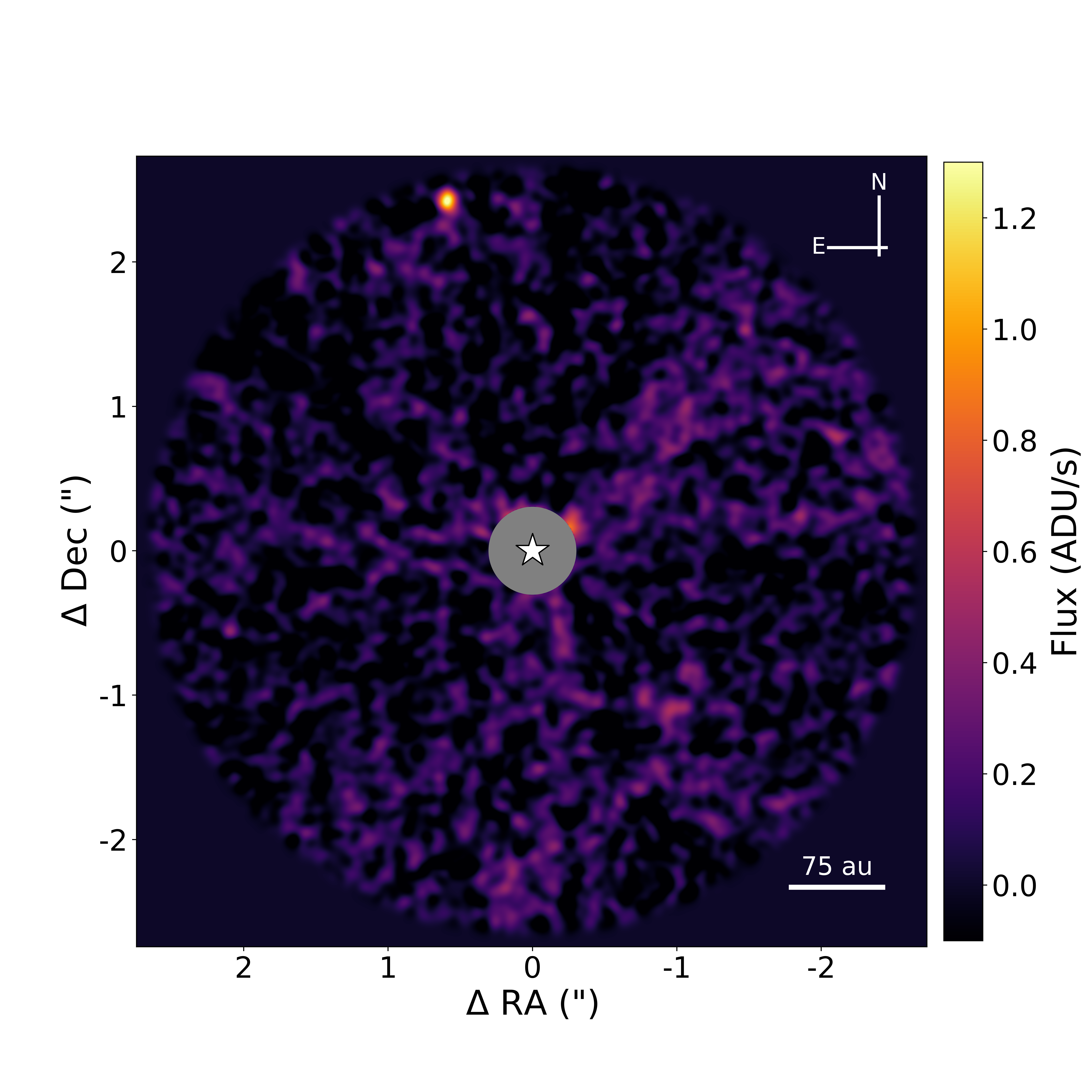}
    \caption{Annular ADI-PCA image of PDS 70 in the M$_s$ band. The background source is visible north of the star. We masked the inner $0.3^{\prime\prime}$ region due to residual speckle noise. }
    \label{pds70}
\end{figure}

PDS 70 is well known to host multiple planets. PDS 70 b, a $7.0\pm2.0$ M\textsubscript{Jup} companion orbiting at $22.7^{+2.0}_{-0.5}$ au, and PDS 70 c, a $4.4\pm1.1$ M\textsubscript{Jup} planet orbiting at $30.2^{+2.0}_{-2.4}$ au \citep{Haffert2019,Mesa2019}. Moreover, two additional candidates have been proposed: PDS 70 d, observed with VLT/SPHERE, VLT/NACO, VLT/SINFONI, and JWST/NIRCam, with a mass of $5.2\pm3.5$ M\textsubscript{Jup} and a semimajor axis of $12.9\pm2.1$ au \citep{Mesa2019,Hammond2025}, and PDS 70 e, inferred from proper motion variation between \textit{Gaia} DR3 and DR2, although its presence remains debated. PDS 70 e is estimated to have a mass of $34^{+300}_{-34}$ M\textsubscript{Jup} at a separation of $4.5^{+20}_{-3.9}$ au \citep{Vioque2026}. 

We detected a point source located north of the star, with a signal-to-noise ratio of $\sim6.8$ (Figure \ref{pds70}). As reported by \citet{Haffert2019}, based on \textit{Gaia} proper motion measurements, and by \citet{Cugno2023}, this source (first reported by \citealt{Riaud2006}, where was mistakenly identified as a bound companion) is a background star. We derive for it a separation of $2.531\pm0.041^{\prime\prime}$ and a position angle of $13.87\pm0.87^\circ$. We do not detect any of the PDS 70 planets, nor any emission from the disk, due to the limited field rotation (Table \ref{obs}), which resulted in poor contrast (Figure \ref{contrasts} and Table \ref{contrast_table}) at all separations.

While the two confirmed planets and PDS 70 d lie below our detection limits, we can place constraints on the mass and separation of PDS 70 e. At separations of $\sim25.3$ au and $\sim12.7$ au, we derive upper mass limits between $\sim7.8$–$9.7$ M\textsubscript{Jup} and $\sim8.3$–$10.1$ M\textsubscript{Jup}, respectively. These values can therefore be considered as upper limits on the mass of PDS 70 e at these separations. Higher masses could still be consistent with closer separations; however, we cannot place constraints in this region, as it partly lies beneath the coronagraph and is partly dominated by residual speckle noise.

\subsection{AS 209}
The presence of a companion orbiting AS 209 remains debated. Based on ALMA $^{13}$CO observations, \citet{Bae2022} reported a $1.3$ M\textsubscript{Jup} candidate at $\sim200$ au, later rejected by \citet{GallowaySprietsma2023} and \citet{Fedele2023}, who instead proposed a $3$–$5$ M\textsubscript{Jup} companion at $\sim100$ au based on disk kinematics. We do not detect any point source in our images. At 200 au, our $5\sigma$ limits ($\sim1.3$–$1.9$ M\textsubscript{Jup}) are comparable to the mass suggested by \citet{Bae2022}, so we cannot rule out such a companion. At 100 au, however, we reach $\sim1.5$–$2.0$ M\textsubscript{Jup}, below the range inferred by \citet{Fedele2023}, and would therefore expect a detection. The non-detection may reflect differences in wavelength sensitivity between our data and ALMA observations, as also suggested by the non-detection in $L'$-band NACO data \citep{Cugno2023}. In addition, the predicted companion properties are strongly model-dependent \citep{Fedele2023}, and the true mass may be lower than estimated.

\bibliography{bibliography}{}
\bibliographystyle{aasjournalv7.1}

\end{document}